      [1999/12/01 v1.4c Il Nuovo Cimento]
\documentclass{cimento}

\usepackage{graphicx} 

\title{Review on Heavy-Ion Physics}
\author{A.~Dainese \from{ins:x}}
\instlist{\inst{ins:x} INFN - Sezione di Padova, via Marzolo 8, 35131 Padova, Italy}
\PACSes{\PACSit{12.38.Mh}{Quark-gluon plasma in quantum chromodynamics} \PACSit{25.75.Nq}{Quark deconfinement, quark-gluon plasma production and phase transitions in relativistic heavy-ion collisions}}

\newcommand{\sqrtsNN}{\sqrt{s_{\scriptscriptstyle{{\rm NN}}}}}

\newcommand{\mev}{\mathrm{MeV}}
\newcommand{\gev}{\mathrm{GeV}}
\newcommand{\tev}{\mathrm{TeV}}
\newcommand{\fm}{\mathrm{fm}}

\newcommand{\PbPb}{\mbox{Pb--Pb}}

\renewcommand{\AA}{\mbox{nucleus--nucleus}}
\newcommand{\RAA}{R_{\rm AA}}

\newcommand{\pt}{p_{\rm t}}

\newcommand{\dNdy}{{\rm d}N_{\rm ch}/{\rm d}y}

\newcommand{\ccbar}{\mbox{$\mathrm {c\overline{c}}$}}

\newcommand{\Jpsi} {\mbox{J\kern-0.05em /\kern-0.05em$\psi$}\xspace}

\let\Otemize =\itemize
\let\Onumerate =\enumerate
\let\Oescription =\description
\def\Nospacing{\itemsep=0pt\topsep=0pt\partopsep=0pt\parskip=0pt\parsep=0pt}
\def\Topspac{\vspace{-0.5\baselineskip}}
\def\Botspac{\vspace{-0.2\baselineskip}}
\newenvironment{Itemize}{\Topspac\Otemize\Nospacing}{\endlist\Botspac}
\newenvironment{Enumerate}{\Topspac\Onumerate\Nospacing}{\endlist\Botspac}

\begin{document}

\maketitle

\begin{abstract}
Collisions of heavy ions (nuclei) at ultra-relativistic energies ($\sqrtsNN\gg~10$~GeV per nucleon--nucleon collision in the centre of mass system) are regarded as a unique tool to produce in the laboratory a high energy density and high temperature state of strongly-interacting  matter. In this short review, we will discuss the 
expected features of this hot and dense state, describe indications on its properties emerged from 
the experimental programs at the CERN-SPS and BNL-RHIC accelerators, and finally 
outlook the perspectives for the forthcoming heavy-ion runs at the CERN-LHC.    
\end{abstract}

\section{Introduction: phenomenology of hot and dense QCD matter}

\begin{figure}[!t]
  \begin{center}
    \includegraphics[width=0.47\textwidth]{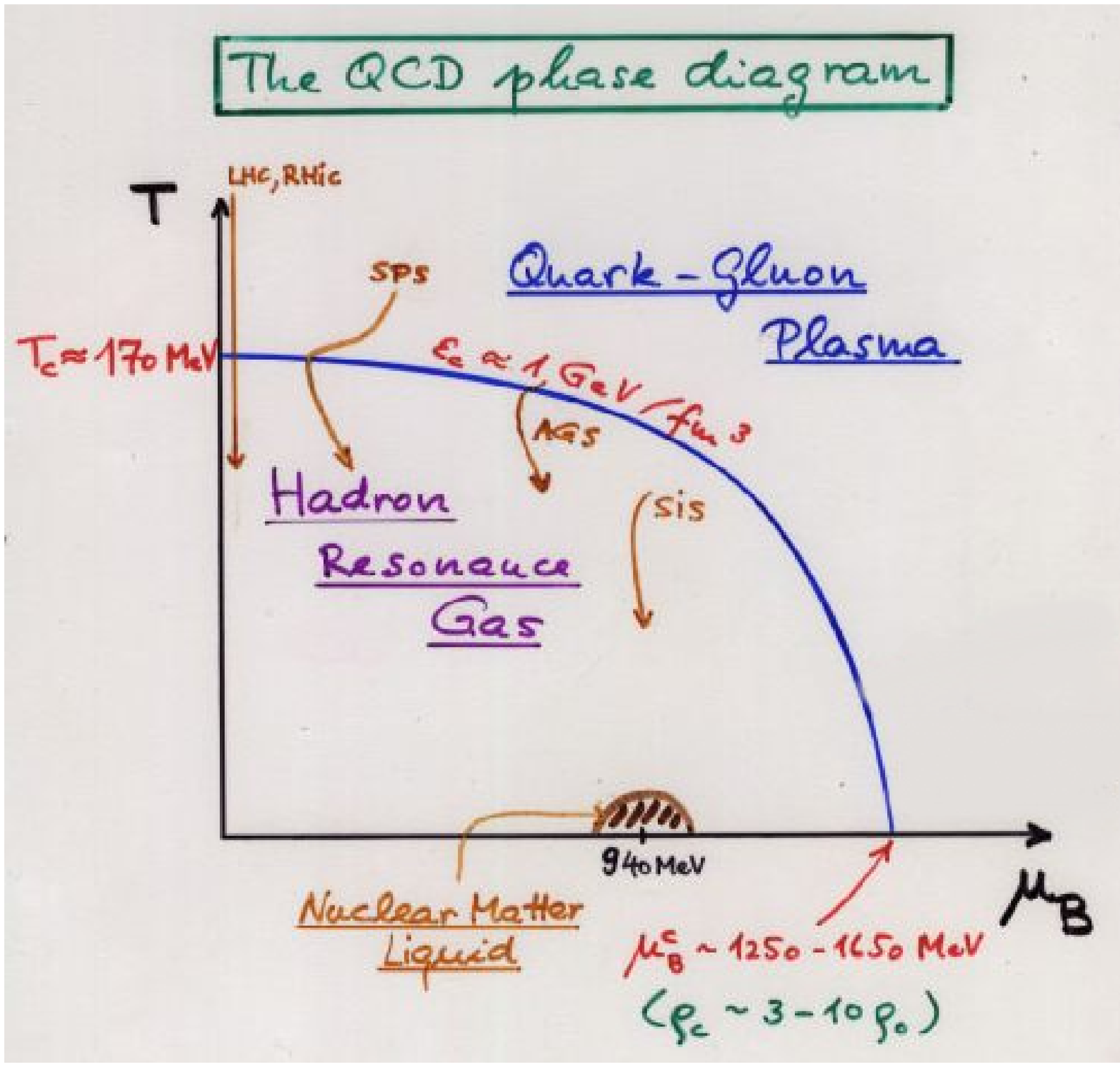}
    \hfill
    \includegraphics[width=0.49\textwidth]{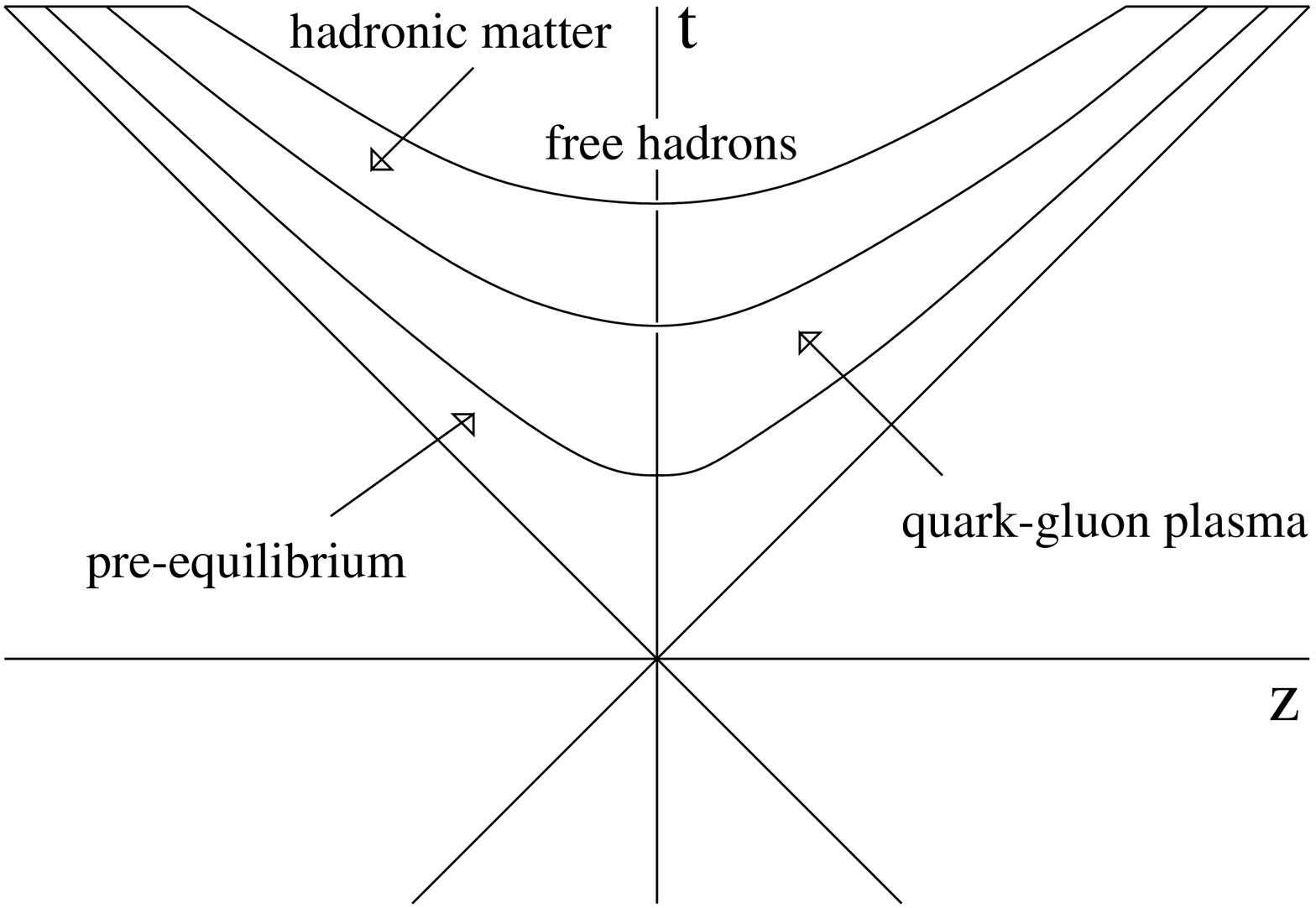}
  \caption{Left: phase diagram of QCD matter~\cite{heinz}.
           Right: the expected evolution of a high-energy nuclear collision.}
  \end{center}
  \label{fig:qcd}
\end{figure}

On the basis of thermodinamical considerations and of QCD calculations, 
strongly interacting matter is expected to exist in different states.
Its behaviour, as a function of the baryonic chemical 
potential
$\mu_{\rm B}$ (a measure of the baryonic density) 
and of the temperature $T$, is displayed in the phase diagram 
reported in Fig.~\ref{fig:qcd} (left). At low temperatures 
and for $\mu_{\rm B}\simeq m_{\rm p}\simeq 940~\mev$, 
we have ordinary matter. When the energy density $\varepsilon$
of the system is increased, by `compression' (towards the right) or by `heating' 
(upward), a phase transition predicted:
the density of partons (quarks and gluons) becomes so high that 
the confinement of coloured quarks in colourless hadrons vanishes and a deconfined Quark-Gluon Plasma
(QGP) is formed. 

From a theoretical point of view, the phase diagram can be explored within the framework of lattice QCD~\cite{wilson,karsch}, although calculations are 
limited to cases of null or small $\mu_{\rm B}$. The phase transition 
is identified by a large increase of the energy density $\varepsilon(T)$ 
when the temperature $T$ reaches the critical value $T_c=(175\pm 15)~\mev$~\cite{karsch}.  
Considering that the energy density of an 
equilibrated ideal gas of particles with $n_{\rm dof}$ degrees of freedom is
$\varepsilon = n_{\rm dof}\,\frac{\pi^2}{30}\,T^4$,
the dramatic increase of $\varepsilon/T^4$ at $T_c$ can be interpreted as due to the 
change of $n_{\rm dof}$ from 3 in the pion gas phase to 37 (with 2 quark flavours)
in the deconfined phase, where the additional colour and quark flavour 
degrees of freedom are available\footnote{In a pion gas the degrees of freedom
are only the 3 values of the isospin for $\pi^+,\,\pi^0,\,\pi^-$. In a
QGP with 2 quark flavours the degrees of freedom are 
$n_g+7/8\,(n_q+n_{\bar{q}})=N_g(8)\,N_{\rm pol}(2)+7/8\,\times 2\times N_{\rm flav}(2)\,N_{\rm col}(3)\,N_{\rm spin}(2)=37$. The factor 7/8 accounts for the 
difference between Bose-Einstein (gluons) and Fermi-Dirac (quarks) 
statistics.}. The energy density for a QGP at $T_c$ is $\varepsilon_c\approx 1~\gev/\fm^3$.

In heavy-ion collisions, both temperature and density increase, possibly 
bringing the system to the phase transition.
In the diagram in Fig.~\ref{fig:qcd} (left) the paths
estimated for the fixed-target (SIS $\sqrtsNN\simeq 1~\gev$, AGS $\sqrtsNN\simeq 5~\gev$, SPS $\sqrtsNN\simeq 17~\gev$) and collider (RHIC $\sqrtsNN=200~\gev$, LHC $\sqrtsNN=2.75$--$5.5~\tev$)
experiments are shown. 
The evolution of a high-energy nucleus--nucleus collision is usually pictured
in the form shown in Fig.~\ref{fig:qcd} (right). 
After a rather short 
equilibration time $\tau_0\simeq 0.1$--$1~\fm/c$, 
the presence of a thermalized medium is assumed, and for sufficiently-high 
energy densities, this medium would be in the Quark-Gluon Plasma phase.
Afterwards, as the expansion reduces the energy density, the system goes 
through a hadron gas phase and finally reaches the freeze-out, when the 
final state hadrons do not interact with each other anymore.

\section{Evidence for deconfinement in heavy-ion collisions: the SPS 
programme}
\label{sec:sps}

The desire to test this fascinating phase structure of 
strongly-interacting matter led, 
in the 1990s, to the fixed-target experiments at the 
AGS in Brookhaven (Au--Au $\sqrtsNN\simeq 5~\gev$) and at the CERN-SPS
(Pb--Pb $\sqrtsNN\simeq 17~\gev$).

The energy density reached in central Pb--Pb collisions at the SPS
was estimated (from the measured transverse energy of the produced 
particles~\cite{bjorkenepsilon,epsilonNA50}) to be $\varepsilon\approx 3~\gev/\fm^3$, 
i.e. above the critical value $\varepsilon_c$.
The two clearest indications for the production of a
deconfined medium in \PbPb~collisions at the SPS are:
\begin{Itemize}
\item the enhancement of the production of strange and multi-strange 
      baryons (hyperons) with respect 
      to the rates extrapolated from pp data (predicted by 
      J.~Rafelski and B.~M\"uller in 1982~\cite{rafelskimuller});
\item the suppression of the production of the J$/\psi$ meson (the lowest 
      $\ccbar$ bound state), always with respect
      to the rates extrapolated from pp (predicted by T.~Matsui and 
      H.~Satz in 1986~\cite{matsuisatz}). 
\end{Itemize}

\begin{figure}[!t]
  \begin{center}
    \includegraphics[width=.7\textwidth]{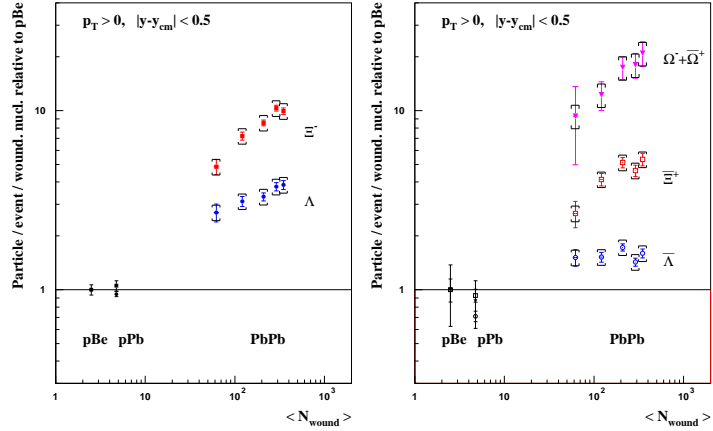}
  \caption{Strange baryon production in \PbPb~per participant nucleon, 
           normalized
           to the ratio from p--Be, as a function of the number of participant
           nucleons, as measured by NA57 at the SPS~\cite{strangenessNA57}.}
  \label{fig:strangeness}
  \end{center}
\end{figure}

In the QGP, the chiral symmetry restoration decreases the threshold 
for the production of a $\rm s\overline{s}$ pair from twice the 
constituent mass of the s quark, $\approx 600~\mev$, to twice the {\sl bare}
mass of the s quark, $\approx 300~\mev$, which is less 
than half of the energy required to produce strange particles 
in hadronic interactions. 
In the QGP, multi-strange baryons can be produced by statistical 
combination of strange (and non-strange) quarks, while in an hadronic gas 
they have to be produced through a chain of interactions that 
increase the strangeness content in steps of one unit. 
For this reason an hyperon enhancement growing with the strangeness content
was indicated as a signal for QGP formation.
This effect was, indeed, observed by the WA97/NA57
experiment~\cite{strangenessNA57}: 
in Fig.~\ref{fig:strangeness} one can see that the production 
of strange and multi-strange baryons increases by 10 times and more
(up to 20 times for the $\Omega$) in central \PbPb~collisions in 
comparison to \mbox{p--Be}, where the QGP is not 
expected. As predicted, the enhancement $\mathcal{E}$ is increasing with 
the strangeness
content: $\mathcal{E}(\Lambda)<\mathcal{E}(\Xi)<\mathcal{E}(\Omega)$.

The other historic predicted signal of deconfinement was  
observed, by the NA50 and NA60 experiments (in Pb--Pb and In--In
collisions, respectively): the suppression 
of the J$/\psi$ particle yield, with respect to the reference Drell-Yan process~\cite{na50jpsi,na60jpsi}.
The suppression is interpreted as due to 
the fact that, in the high colour-charge density environment of a QGP,   
the strong interaction between the two quarks of the $\ccbar$ pair is 
screened and the formation of their bound state is consequently prevented.

\section{RHIC: focus on new observables}
\label{rhic}

The Relativistic Heavy Ion Collider (RHIC) in Brookhaven began 
operation during summer 2000. With a factor 10 increase in the centre-of-mass
energy with respect to the SPS, $\sqrtsNN$ up to $200~\gev$, 
the produced collisions are expected to be well above the phase transition 
threshold. Moreover, in this energy regime, the so-called 
`hard processes' ---production 
of energetic partons ($E>3$-$5~\gev$) out of the inelastic scattering of two 
partons from the colliding nuclei--- have a significantly large cross section 
and they become experimentally accessible. 

In this scenario, beyond the `traditional' observables
we have already introduced, the interesting phenomenon of in-medium
parton energy loss~\cite{bjorkenjets,gyulassywang,bdmps}, predicted 
for the first time by J.D.~Bjorken in 1982~\cite{bjorkenjets}, 
can be addressed. 

\begin{figure}[!t]
  \begin{center}
  \includegraphics[width=.49\textwidth]{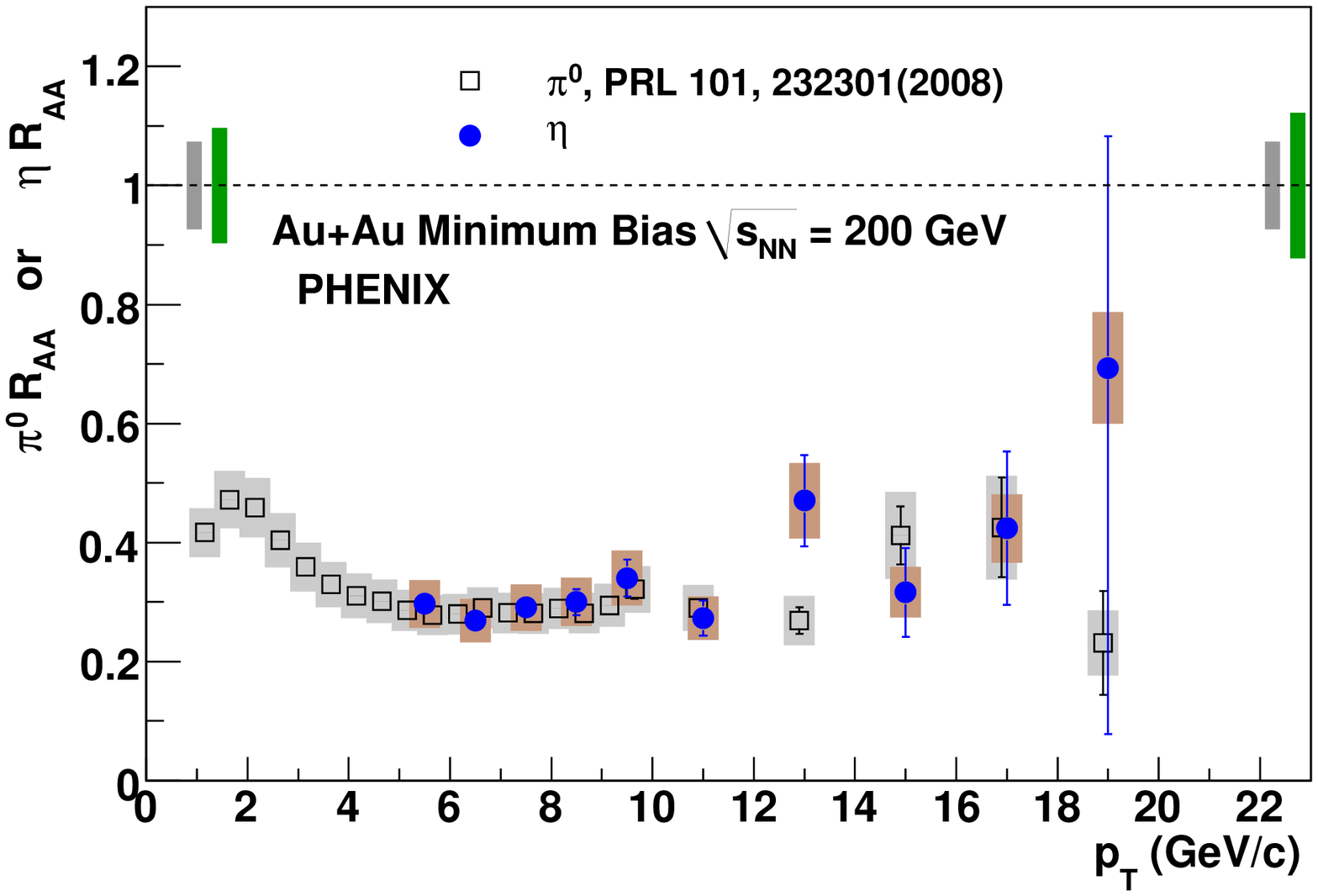}
  \includegraphics[width=.49\textwidth]{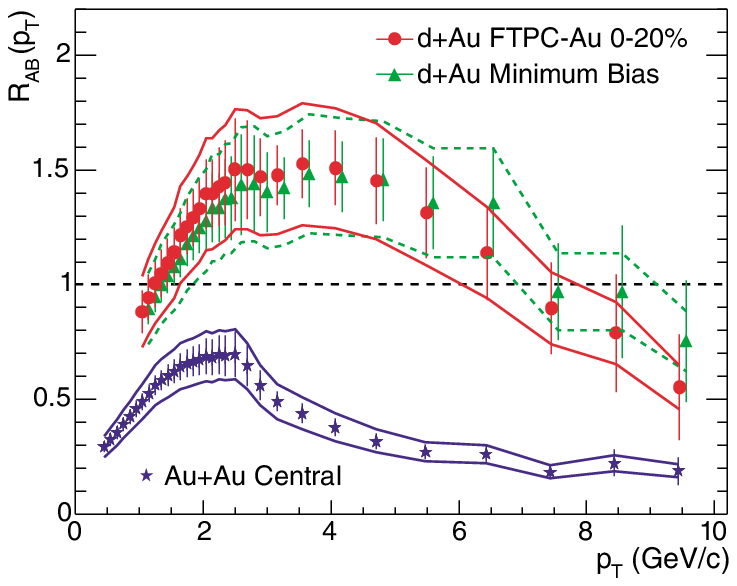}
  \caption{The ratio of transverse momentum distributions of 
           hadrons in Au--Au collisions and pp 
           collisions, scaled by the number of binary nucleon--nucleon 
           collisions at $\sqrtsNN=200~\gev$. Left: PHENIX data for $\pi^0$ and 
           $\eta$~\cite{phenixRAA}. Right: STAR data for charged 
           hadrons~\cite{starRAAandAzi}; also 
           the ratio d--Au to pp is shown.}
  \label{fig:RAA}
   \vglue0.3cm
  \includegraphics[width=.6\textwidth]{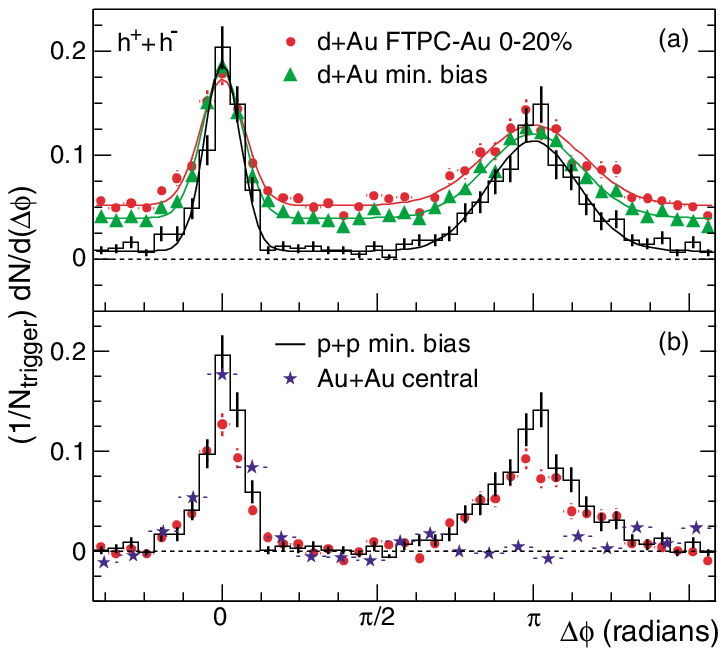}
  \caption{Azimuthal correlations of charged particles relative to a 
           high-$\pt$ trigger particle for  
           central Au--Au, compared to d--Au and pp collisions at 
           $\sqrtsNN=200~\gev$~\cite{starRAAandAzi}.}
  \label{fig:starAzi}
  \end{center}
\end{figure}

Hard partons are produced at the early stage of the collision and they 
propagate through the medium formed in the collision. During this 
propagation they undergo interactions (both inelastic and elastic) 
with the gluons present in the 
medium and they lose energy. Such energy loss is not peculiar of a 
deconfined medium, but, quantitatively, it is strongly dependent on the 
nature and on the properties of the medium, being predicted to be much 
larger in the case of deconfinement. 

The measurement of high-$\pt$ (projection of the momentum on the plane 
transverse to the beam line) particle production has been addressed at RHIC mainly 
by the PHENIX~\cite{phenix} and STAR~\cite{star} experiments, but has been observed 
also by the two smaller experiments BRAHMS~\cite{brahms} and PHOBOS~\cite{phobos}. 
The results are striking.
Figure~\ref{fig:RAA} reports the yield of neutral and charged hadrons measured 
by PHENIX~\cite{phenixRAA} and STAR~\cite{starRAAandAzi} in  
central Au--Au collisions at 
$\sqrtsNN=200~\gev$, divided by the yield in pp collisions
at the same energy and by the estimated number of binary 
nucleon--nucleon collisions. 
This ratio, $\RAA$, should be one at high $\pt$ if no medium
effects are present. In central 
collisions the yield is reduced of a factor 4--5 
with respect to what expected for incoherent
production in nucleon--nucleon collisions. 

Another impressive result, obtained by STAR and PHENIX, is the 
disappearance of the back-to-back azimuthal correlations of high-$\pt$ 
particles in central Au--Au collisions. 
In Fig.~\ref{fig:starAzi}
the azimuthal correlations of charged particles with respect to 
a high-$\pt$ trigger particle (star markers) in central Au--Au
are shown 
and compared with reference data from pp collisions (histogram):
in central collisions the opposite-side ($\Delta\phi=\pm \pi$)
correlation is strongly suppressed with respect to the pp
and peripheral Au--Au cases. This effect suggests 
the absorption of one of the two jets (usually produced as 
back-to-back pairs) in the hot matter formed in central collisions.

The effects of leading particle and jet suppression are not observed 
in d--Au (deuteron--gold) collisions at 
$\sqrtsNN=200~\gev$ (circle and triangle markes in 
Figs.~\ref{fig:RAA}-right and~\ref{fig:starAzi}), 
where the formation of a dense medium is not expected. 

In summary, experiments at RHIC have shown that a very dense QCD medium 
is formed in high-energy heavy-ion collisions. Other measurements, 
namely elliptic flow and baryon-to-meson ratios, indicate that this medium 
is characterized by partonic degrees of freedom and that its expansion and 
cooling is well described by hydrodynamical models with high viscosity.
Thus, this medium is more similar to a liquid than to a gas of gluons and quarks.

\section{LHC: study of `deeply deconfined' matter}
\label{CHAP1:lhc}

The Large Hadron Collider (LHC) started operation with pp collisons
at $\sqrt{s}=900~\gev$ (2009) and 7~TeV (since March 2010). 
The first Pb--Pb run, at $\sqrtsNN=2.75~\tev$, is scheduled for November 2010.
The maximum Pb--Pb energy of 5.5~TeV should be reached in 2013. 
LHC will provide nuclear collisions at a centre-of-mass energy 15--30 times 
higher than at RHIC, opening a new era for the field, in which particle 
production will be dominated by hard processes, and the energy densities 
will possibly be high enough to treat the generated Quark-Gluon Plasma
as an ideal gas. These qualitatively new features will allow to address
the task of the LHC heavy ion program: a {\sl systematic study of the 
properties of the Quark-Gluon Plasma state. }

Table~\ref{tab:spsrhiclhc} presents a comparison 
of expected values for 
the most relevant parameters for SPS, RHIC and LHC energies~\cite{eskola1}.
At the LHC, the high energy in the collision centre of mass is expected to 
determine a large energy density and an initial temperature at least a factor 
2 larger than at RHIC. This high initial temperature extends also the 
life-time and the volume of the deconfined medium, since it has to expand 
while cooling down to the freeze-out temperature, which is $\approx 170~\mev$
(it is independent of $\sqrt{s}$, above the SPS energy). In addition, 
the large expected number of gluons favours energy and momentum exchanges, 
thus considerably reducing the time needed for the thermal equilibration
of the medium. To summarize, the LHC will produce {\sl hotter, 
larger and longer-living} `drops' of QCD plasma than the previous 
heavy-ion facilities.

The key advantage in this new `deep deconfinement' scenario is that the 
Quark-Gluon Plasma studied by the LHC experiments 
will be much more similar to the Quark-Gluon Plasma that can be investigated 
from a theoretical point of view by means of lattice QCD. 

\begin{table}[!t]
\caption{Comparison of the parameters characterizing central \AA~collisions
         at different energy regimes~\cite{eskola1}.}
\label{tab:spsrhiclhc}
\begin{center}
\begin{tabular}{ll|ccc}
\hline
Parameter & & SPS & RHIC & LHC \\
\hline
$\sqrtsNN$ & [GeV] & 17 & 200 & 2750--5500 \\
d$N_{\rm gluons}/$d$y$ & & $\simeq 450$ & $\simeq 1200$ & $\simeq 5000$ \\
$\dNdy$ & & 400 & 650 & $\simeq 2000$--3000 \\
Initial temperature & [MeV] & 200 & 350 & $>600$ \\
Energy density & [$\gev/\fm^3$] & 3 & 25 & 120 \\
Freeze-out volume & [$\fm^3$] & few $10^3$ & few $10^4$ & few $10^5$ \\
Life-time & [$\fm/c$] & $<2$ & $2$-$4$ & $>10$ \\ 
\hline
\end{tabular}
\end{center}
\end{table}

Heavy-ion collisions at the LHC access not only a quantitatively different
regime of much higher energy density but also a qualitatively new regime, 
mainly because:
\begin{Enumerate}
\item {\sl High-density parton distributions } are expected to dominate 
      particle production.
      The number of low-energy partons (mainly gluons) in the two colliding
      nuclei is, therefore, expected to be so large as to produce a 
      significant initial-state saturation that would be the dominant effect 
      governing the bulk particle production.
\item {\sl Hard processes } should
      contribute significantly to the total nucleus--nucleus cross section.
      The hard probes are at the LHC an ideal experimental tool for a 
      detailed characterization of the QGP medium.
\end{Enumerate}

Three experiments will study Pb--Pb collisions at the LHC: ALICE (the dedicated
heavy-ion experiment), ATLAS, and CMS. Their excellent detector capabilities,
with many complementary features, will allow to pursue 
an exciting heavy-ion physics program. As aforementioned, the goal of 
this program is a systematic study of the properties of the Quark-Gluon Plasma
state. In detail:

\begin{Itemize}
\item Energy density of the hot and dense system produced in the collision,
via measurement of the charged-particle multiplicity and charged-particle rapidity density.
\item Temperature and baryon density of the system at the chemical freeze-out and temperature at the kinetic freeze-out, via
 measurement of identified-particle yields (probe chemical freeze-out) and momentum spectra (probe kinetic freeze-out).
\item Hadronization mechanism of the system: interplay between quark recombination from a partonic medium and parton fragmentation outside the medium,
via measurement of the baryon-to-meson ratios as a function of momentum and rapidity.
\item Size of the hot particle-emitting source, via Bose-Einstein interferometry with identical bosons.
\item Pressure-driven expansion of the system, to be compared to hydrodynamical models, via measurement of radial flow and elliptic flow (azimuthal particle production anisotropy in non-central collisions).
\item Fluctuations induced by the QCD phase transition, via measurement of event-by-event particle spectra
\item Effect of parton energy loss  on heavy quarks (charm and beauty): dependence of energy loss on the parton colour charge (c quarks vs. gluons) and on the parton mass (b quarks vs. c quarks),
via measurement of transverse momentum spectra of D and B mesons.
\item Effect of parton energy loss on jet structure and jet fragmentation function, via measurement of reconstructed jets and photon-tagged jets (provide initial jet energy calibration).
\item Quarkonium states (J$/\psi$, $\psi^\prime$, $\Upsilon$, $\Upsilon^\prime$, $\Upsilon^{\prime\prime}$) suppression pattern 
by temperature-dependent dissociation due to colour screening 
in a deconfined medium, via measurement of charmonium and bottomonium yields.
\item Initial temperature of the QGP (which would radiate thermal photons), via measurement of the transverse momentum spectrum of single photons.
\end{Itemize}

\section{Summary}

High-energy heavy-ion physics addresses the limits of colour confinement and
the properties of QCD in extended high-density systems.

Experiments at the SPS and RHIC accelerators have provided indications on the 
properties of the medium formed in the collisions: the medium is deconfined, 
partonic, dense, and hydrodynamic.  

The LHC machine and experiments will bring a big step forward in the exploration
of these new territories of QCD, because the medium will be hotter and denser,
because the probes for its characterization will be abundantly available, and 
because the new generation detectors will provide unprecedented experimental performance.

\acknowledgments
The author thanks the Organizers of the IFAE2010 Conference 
for the kind invitation.

\end{document}